\documentclass[aps,prc,groupedaddress,reprint,showpacs,floatfix]{revtex4-2}
\usepackage{graphicx}
\begin{document}
\title{To examine the variation in dissipation near the shell closure using neutron multiplicity as a probe}
\author{Punit Dubey} \affiliation{Department of Physics, Banaras Hindu University, Varanasi-221005, India}
\author{Mahima Upadhyay} \affiliation{Department of Physics, Banaras Hindu University, Varanasi-221005, India}
\author{Mahesh Choudhary} \affiliation{Department of Physics, Banaras Hindu University, Varanasi-221005, India}
\author{Namrata Singh} \affiliation{Department of Physics, Banaras Hindu University, Varanasi-221005, India}
\author{Shweta Singh} \affiliation{Department of Physics, Banaras Hindu University, Varanasi-221005, India}
\author{N. Saneesh} \affiliation{Inter-University Accelerator Centre, Aruna Asaf Ali Marg, New Delhi - 110067, India}
\author{Mohit Kumar} \affiliation{Inter-University Accelerator Centre, Aruna Asaf Ali Marg, New Delhi - 110067, India}
\author{Rishabh Prajapati} \affiliation{Inter-University Accelerator Centre, Aruna Asaf Ali Marg, New Delhi - 110067, India}
\author{K. S. Golda} \affiliation{Inter-University Accelerator Centre, Aruna Asaf Ali Marg, New Delhi - 110067, India}
\author{Akhil Jhingan} \affiliation{Inter-University Accelerator Centre, Aruna Asaf Ali Marg, New Delhi - 110067, India}
\author{P. Sugathan} \affiliation{Inter-University Accelerator Centre, Aruna Asaf Ali Marg, New Delhi - 110067, India}
\author{Jhilam Sadhukhan} \affiliation{Variable Energy Cyclotron Centre, Kolkata-700064, India}
\author{Raghav Aggrawal} \affiliation{Department of Physics, Panjab University, Chandigarh-160014, India}
\author{Kiran} \affiliation{Department of Physics, Panjab University, Chandigarh-160014, India}
\author{A. Kumar}\email{ajaytyagi@bhu.ac.in}
\affiliation{Department of Physics, Banaras Hindu University, Varanasi-221005, India}
\date{\today}
\begin{abstract}
 The pre and post-scission neutron multiplicities have been determined for the fission of the compound nucleus (CN) $^{206}$Rn, induced by the reaction $^{28}$Si+$^{178}$Hf within the excitation energy interval of 61.0-90.0 MeV. We intentionally formed CN $^{206}$ Rn, which is below the shell closure CN, to examine the variation in N/Z with total neutron multiplicity, as data for other CNs of $^{208,210,212,214,216}$ Rn have already been published in the literature. We identified a new trend in the N/Z ratio, where the total neutron multiplicity initially decreases as we approach the shell closure of the compound nucleus and then starts to increase as we move away from the shell closure. Furthermore, we have observed that below the neutron shell closure, the dissipation in compound nuclei (CN) escalates with rising excitation energy, remains stable at the shell closure CN, and thereafter diminishes with increasing excitation energy above the shell closure CN.
\end{abstract}
\maketitle
\section{Introduction}
To understand nuclear dynamics in low-energy heavy-ion collisions, different experimental probes are used, such as measuring the mass, total kinetic energy (TKE), particle multiplicities and angular distribution of fission fragments, and their correlations. The pre-scission neutron multiplicity (M$_{pre}$) is now recognized as one of the most efficient methods for examining the fission time scale in heavy-ion-induced fusion-fission (HIFF) reactions \cite{01}. This is achievable because the neutrons released from a compound nucleus prior to fission are kinematically distinct from those emitted by the fission fragments, allowing for precise measurement of their multiplicity. The observed multiplicities of the pre-scission neutrons were notably higher than what the statistical model of compound nucleus decay had anticipated \cite{01, 02}. Similar observations were noted for pre-scission light-charged particles \cite{03} and Giant dipole resonance (GDR) gamma ($\gamma$) rays \cite{04}. Numerous experimentalists have determined the neutron multiplicity utilizing many different types of projectile target combinations \cite{05, 06, 07, 08, 09, 10, 11, 12, 13, 14, 15, 16, 17, 18, 19, 20, 21, 22, 23, 24} to elucidate the effects of the entrance channel, shell closure, neutron-to-proton ratio (N/Z), excitation energy, and few other physical phenomena.\\
Nuclear dissipation plays a critical role in HIFF dynamics. Several studies in the different mass regions have shown an effect of dissipation in nuclear dynamics \cite{09, 10, 11, 12, 13, 16, 25, 26, 27, 28}. The influence of entrance channels on the dynamics of HIFF and the formation of the compound nucleus through various combinations has been documented in several studies \cite{11, 12}. The entrance channel mass asymmetry, defined as $\alpha$ = (A$_{t}$ - A$_{p}$)/(A$_{t}$ + A$_{p}$), is recognized as a significant factor affecting the dynamical evolution of a dinuclear system, which ultimately leads to the formation of a compound nucleus. The fusion process follows different paths depending on whether $\alpha$ is smaller or larger than a critical value called the Businaro-Gallone mass asymmetry ($\alpha_{BG}$) \cite{29, 30}. Experiments \cite{11, 12} have shown that when $\alpha$ is less than $\alpha_{BG}$, the number of pre-scission (M$_{pre}$) is higher than when $\alpha$ is greater than $\alpha_{BG}$, highlighting the impact of the effects of the entrance channel on fission dynamics.
\\
Further investigation noted that dissipation is comparatively weaker in a shell-closed nucleus compared to neighboring nuclei that do not exhibit shell closure \cite{07, 31}. Some studies also indicate that as the excitation energy increases, the shell effect in M$_{pre}$ transitions from a strength state to a frail state \cite{09, 31}. Only few experimental and theoretical studies have been conducted in the past to examine the influence of N/Z on the pre-scission neutron multiplicity. W. Ye \cite{32} has observed that as the N/Z of the system increases, M$_{pre}$ also increases, while Sandal et al. \cite{09} indicated that the dissipation strength in relation to N/Z does not exhibit any particular trend.\\
To investigate the N/Z dependency in a broad isotopic range, we conducted experimental measurements of the neutron multiplicities for compound nuclei $^{206}$Rn, generated by the reaction of $^{28}$Si with $^{178}$Hf. Furthermore, we incorporated existing experimental data for compound nuclei $^{208,210,212,214,216}$Rn \cite{09, 10} from EXFOR. In this study, we initially correlated our experimental findings with the shell closure systematic along with existing experimental data on neutron multiplicities. Furthermore, we compared our findings with the dynamical model code VECLAN and the systematic equations. In addition, we also calculated the dissipation strength and fission times for the present measurements and correlated them with the existing data in the literature. \\
The current work is structured into five distinct sections. Section II provides the experimental details, while Section III presents the data analysis and results. Discussions will be found in Section IV and the outcome of our study is described in Section V.
\section{Experimental Details} 
The experiment was carried out at the National Array of Neutron Detectors (NAND) facility of the Inter-University Accelerator Center (IUAC) in New Delhi, using a pulsed beam of $^{28}$Si with a repetition rate of 250 ns, produced by the 15UD Pelletron + LINAC accelerator, which was directed at the target $^{178}$Hf with a thickness of 350 $\mu$g/cm$^{2}$. The target was prepared using the ultra-high vacuum evaporation technique \cite{33} with a carbon backing of 40 $\mu$g/cm$^{2}$. The target was positioned in the middle of a spherical chamber with a diameter of 100 cm. Measurements were conducted at four distinct excitation energies of the CN: E{$^{*}$}$_{CN}$ = 61.0, 71.7, 79.0, and 90.0 MeV. The entire experimental setup is shown in Fig. 1.
\begin{figure}
\begin{center}
\includegraphics[width=9.0 cm] {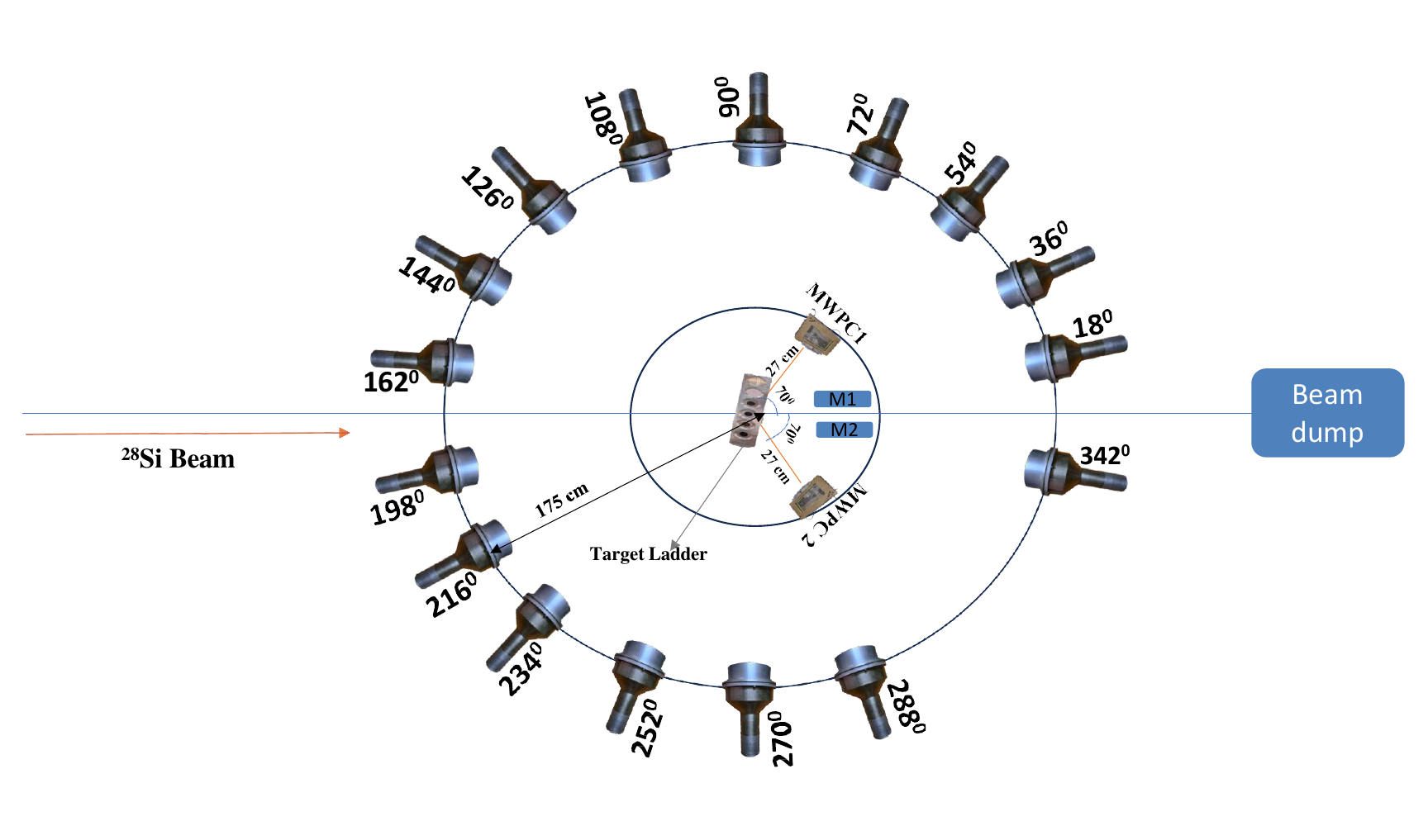}
\caption{Experimental setup (top view) used for the measurement of neutron multiplicity in coincidence with fission fragments for $^{28}$Si+$^{178}$Hf reaction.}
\end{center}
\end{figure} 
\begin{figure}
\begin{center}
\includegraphics[width=9.0 cm] {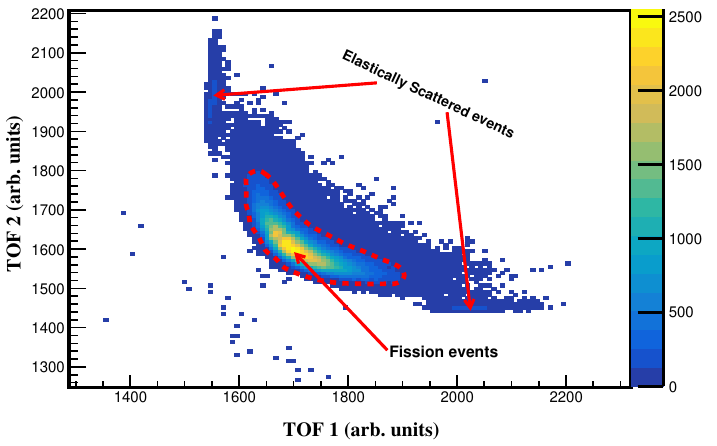}
\caption{Time correlation spectra of complementary fission fragment detected in the MWPCs for $^{28}$Si+$^{178}$Hf at E$^{*}$=79.0 MeV.}
\end{center}
\end{figure} 
\begin{figure}
\begin{center}
\includegraphics[width=9.0 cm] {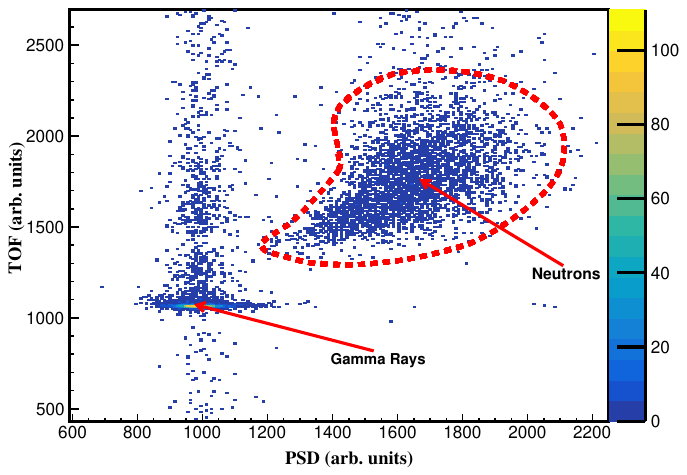}
\caption{Pulse Shape Discrimination (PSD) vs Time of flight (TOF) spectrum of a neutron detector for $^{28}$Si+$^{178}$Hf at E$^{*}$=79.0 MeV.}
\end{center}
\end{figure} 

\begin{figure*}
\begin{center}
\includegraphics[width=18.0 cm, height=10.0 cm]{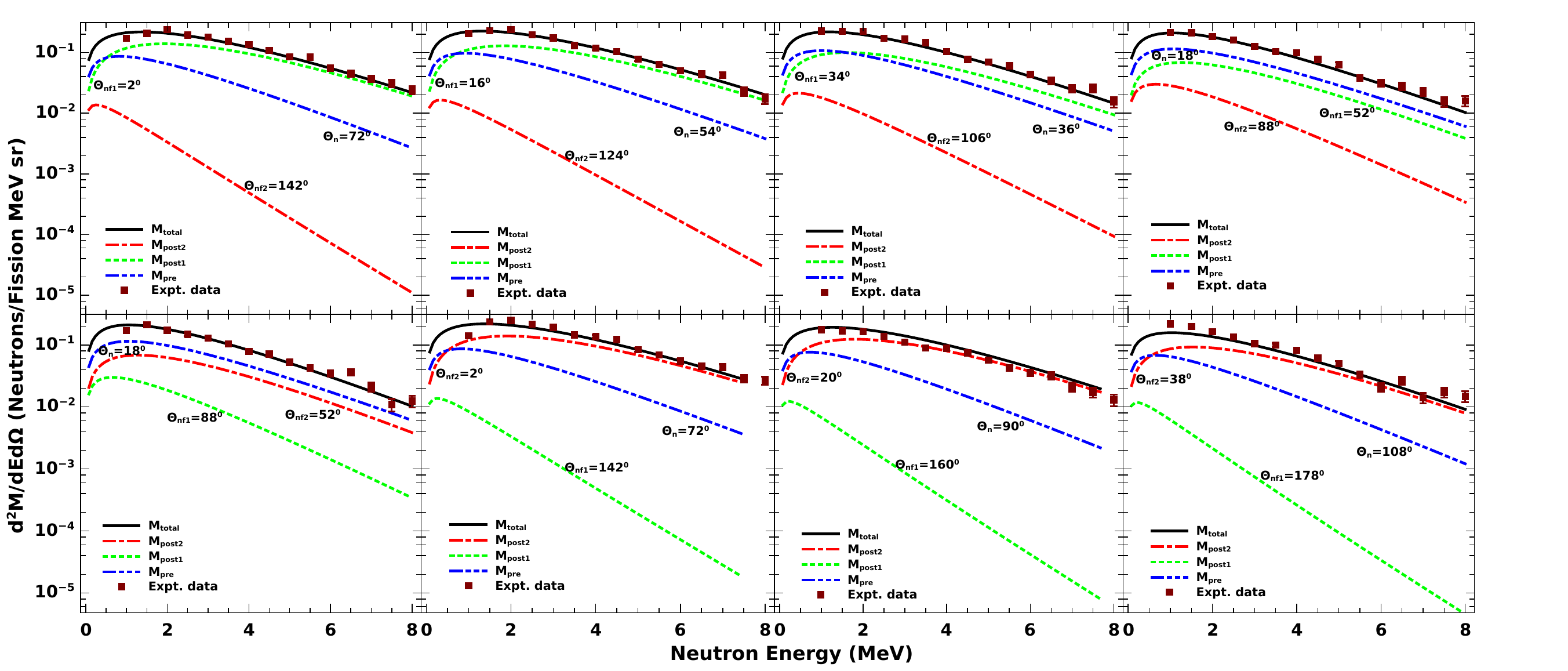}
\caption{Double differential neutron multiplicity spectra for the reaction $^{28}$Si+$^{178}$Hf at E$^{*}$=79.0 MeV, $\theta$$_{n}$ represent polar angle of neutron detector, $\theta$$_{nf1}$ and $\theta$$_{nf2}$ are angles between neutron detector and fission fragments. Square represents present data, M$_{pre}$ (dot-dot-dash), M$_{total}$ (solid line), M$_{post1}$ (dotted line), M$_{post2}$ (dot-dashed) lines from fragments are obtained by fitting experimental values in Eq. (1).}
\end{center}
\end{figure*}
Two multiwire proportional counters (MWPCs) with an active area of 20 × 10 cm² \cite{34} each were employed to detect the complementary fission fragments. The MWPCs were positioned at an angle of 70$^{0}$ relative to the beam direction (both sides) and at distances of 27 cm from the target, corresponding to the folding angle between the fragments. These MWPCs were operated using isobutane gas at a pressure of 4 mbar. To monitor the beam, two silicon surface barrier detectors were also placed within the chamber at angles of $\pm$12$^{0}$ with respect to the direction of the beam.\\
An array of 16 organic liquid scintillator detectors (BC501) were used to detect neutrons emitted from the CN and fission fragments \cite{35}. The flight path for all the neutrons that were detected is 175 cm in length. The detectors were placed in a circle around the reaction plane to achieve angular coverage ranging from 18$^{0}$ to 342$^{0}$. The threshold voltage for the neutron detectors was maintained at around 0.5 MeV by calibration using standard $\gamma$-ray sources ($^{137}$Cs and $^{60}$Co) \cite{36}. Background neutrons were mitigated by positioning a beam dump 4.5 meters downstream of the target position, along with appropriate paraffin and lead brick shielding. Data were collected in event mode using the ROOT-based data acquisition system \cite{37}. 
\section{Data Analysis and Results}
\begin{table*}
\caption{\label{tab:table1}Experimentally calculated results for $^{28}$Si+$^{178}$Hf reaction forming $^{206}$Rn$^{*}$ at various excitation energies.}
\begin{ruledtabular}
\begin{tabular}{cccccc}
 E$^*$ (MeV) &$M_{pre}$ & $2M_{post}$ & $M_{total}$ &$T_{pre}$ & $T_{post}$ \\ \hline
 61.00 & 2.12$\pm$0.13 & 3.26$\pm$0.07 & 5.38$\pm$0.14 & 1.34$\pm$0.07 & 1.11$\pm$0.03  \\
71.70 & 2.79$\pm$0.13 & 3.46$\pm$0.07 & 6.25$\pm$0.15 & 1.46$\pm$0.06 & 1.17$\pm$0.03  \\
79.00 & 3.15$\pm$0.16 & 3.74$\pm$0.10 & 6.89$\pm$0.19 & 1.47$\pm$0.06 & 1.26$\pm$0.03  \\
90.00 & 3.32$\pm$0.18 & 3.84$\pm$0.11 & 7.15$\pm$0.21 & 1.56$\pm$0.07 & 1.28$\pm$0.04  \\
\end{tabular}
\end{ruledtabular}
\end{table*}

The binary fission data was analyzed using the ROOT-based analysis framework. Fission fragments were separated from other charged particles such as scattered projectiles, target recoils, etc., by time of flight (TOF) and kinematic coincidence. Fig. 2 shows the time correlation of events detected in both the MWPCs. Excellent neutron-$\gamma$ separation was obtained from TOF and zero-crossing time measurements. Fig. 3 shows a two-dimensional (2D) histogram of the TOF versus pulse shape discrimination (PSD) from one of the neutron detectors. It can be seen that the neutron events are clearly separated from the $\gamma$-ray background. Applying a software cut around neutron events [shown by the dashed loop in Fig. 3], the neutron TOF was converted to an energy histogram, considering the position of the prompt $\gamma$ peak as the time reference. An energy-dependent efficiency correction was applied using the measured intrinsic efficiency of the neutron detector in the energy range of 1 to 8 MeV. \\
The kinematic effects on the energy and angular distributions of neutrons are assumed to originate from three moving sources (CN and two fission fragments), and the corresponding neutron spectra are described by the Watt distributions \cite{38},
$$ \frac {d^{2}M} {dE_{n}d\Omega_{n}}= \sum_{i=1}^{3} \frac{M_{i}\sqrt{E_{n}}}{2({\pi}{T_{i}})^{3/2}} $$
 \begin{equation} \times exp[-{\frac{E_{n}-2\sqrt{E_{n}E{i}/A_{i}}cos{\theta}_{i}+E_{i}/A_{i}}{T_{i}}}],
 \end{equation} 
where A$_{i}$, E$_{i}$, T$_{i}$, and M$_{i}$ are the mass number, kinetic energy, temperature, and multiplicity of each neutron emitting source i, respectively. E$_{n}$ is the laboratory energy of the neutron and d$\Omega_{n}$ is the solid angle subtended by each BC501A detector. $\theta$$_{i}$ is the relative angle between the neutron source and the neutron detector. $\theta$$_{n}$, $\theta$$_{nf1}$, and $\theta$$_{nf2}$ shown in Fig. 4 represent the angles between the direction of neutron and neutron emitting sources such as CN, fragment 1 (f$_{1}$), and fragment 2 (f$_{2}$), respectively. The temperature of the fissioning nucleus ($T_{pre}$) is determined using the formula \cite{39}, $T_{pre}$ = 11/12$\sqrt{E^{*}/a}$, where $E^{*}$ represents the excitation energy of the CN, and 'a' is the level density parameter, defined as A$_{CN}$/9 MeV$^{-1}$ \cite{06}. Because only the symmetric fission mode is considered, both fragments have identical values for M$_{post}$ and T$_{post}$. The total neutron multiplicity is thus obtained as M$_{total}$ = M$_{pre}$+2M$_{post}$. To extract M$_{pre}$ and M$_{post}$, a global fit to experimental spectra of d$^{2}$M/$dE_{n}d\Omega_{n}$ was made in terms of the Watt expression. The method used to extract pre-scission neutron multiplicity is based on fitting the neutron energy spectra at different angles using $\chi$$^{2}$ minimization. The energy of the fission fragments and the folding angles were calculated using the Viola systematics for symmetric fission \cite{40}. To reduce any uncertainty in angles due to the large area of fission detectors, the data were analyzed only when the detected fragment was located within a rectangular slice of MWPC covering $\pm$10$^{0}$. Fig. 4 shows an example of the moving source fit to the experimental neutron multiplicity spectrum in the laboratory frame for E$^{*}$ = 79.0 MeV. In Fig. 4, the double differentials of neutron multiplicity are shown as a function of neutron energy for eight NAND detectors near the MWPCs in the reaction plane. As expected, due to kinematic focusing, these spectra are dominated by contributions from their respective fission fragments. This spectrum shows the largest contribution from pre-scission neutrons (CN source). The excellent agreement between multiple moving-source fits and experimental spectra is displayed in Fig. 4 indicates the data are well described by three moving sources. The average M$_{pre}$, M$_{post}$, M$_{total}$, T$_{pre}$, and T$_{post}$, extracted from the fitted spectra are listed in Table 1.
\section{Discussions}
\subsection{Systematics of pre-scission neutron multiplicity}
\begin{table*}
\caption{\label{tab:table2}Details of the reactions selected in this work. The $\delta_{sh}$ is defined in the text.}
\begin{ruledtabular}
\begin{tabular}{cccccccccc}
 S. No. & CN & Reaction & $\delta_{sh}$ & Ref. & S. No. & CN & Reaction & $\delta_{sh}$ & Ref. \\ \hline
1a & $^{243}$Am & $^{11}$B+$^{232}$Th  & 19 & \cite{05} & 5b & $^{200}$Pb & $^{19}$F+$^{181}$Ta  & -24 & \cite{08}  \\
1b & $^{248}$Cf & $^{11}$B+$^{237}$Np  & 24 & \cite{05} & 5c & $^{203}$Bi & $^{19}$F+$^{184}$W  & -21 & \cite{22}  \\
2a & $^{206}$Po & $^{12}$C+$^{194}$Pt  & -18 & \cite{06} & 5d & $^{213}$Fr & $^{19}$F+$^{194}$Pt  & -11 & \cite{07}  \\
2b & $^{210}$Po & $^{12}$C+$^{198}$Pt  & -14 & \cite{06} & 5e & $^{215}$Fr & $^{19}$F+$^{196}$Pt  & -9 & \cite{07}  \\
2c & $^{216}$Ra & $^{12}$C+$^{204}$Pb  & -8 & \cite{11} & 5f & $^{216}$Ra & $^{19}$F+$^{197}$Au  & -8 & \cite{11}  \\
2d & $^{244}$Cm & $^{12}$C+$^{232}$Th  & 20 & \cite{05} & 5g & $^{217}$Fr & $^{19}$F+$^{198}$Pt  & -7 & \cite{07}  \\
3a & $^{197}$Tl & $^{16}$O+$^{181}$Ta  & -27 & \cite{12} & 5h & $^{228}$U & $^{19}$F+$^{209}$Bi  & 4 & \cite{13}  \\
3b & $^{210}$Rn & $^{16}$O+$^{194}$Pt  & -14 & \cite{09} & 5i & $^{251}$Es & $^{19}$F+$^{232}$Th  & 27 & \cite{08}  \\
3c & $^{213}$Fr & $^{16}$O+$^{197}$Au  & -11 & \cite{08} & 6a & $^{229}$Np & $^{20}$Ne+$^{209}$Bi  & 5 & \cite{17}  \\
3d & $^{214}$Rn & $^{16}$O+$^{198}$Pt  & -10 & \cite{09} & 7a & $^{198}$Pb & $^{28}$Si+$^{170}$Er  & -24 & \cite{18}  \\
3e & $^{220}$Th & $^{16}$O+$^{204}$Pb  & -4 & \cite{14} & 7b & $^{206}$Rn & $^{28}$Si+$^{178}$Hf  & -18 & present work  \\
3f & $^{222}$Th & $^{16}$O+$^{206}$Pb  & -2 & \cite{14} & 8a & $^{200}$Pb & $^{30}$Si+$^{170}$Er  & -22 & \cite{18}  \\
3g & $^{224}$Th & $^{16}$O+$^{208}$Pb  & 0 & \cite{14} & 8b & $^{208}$Rn & $^{30}$Si+$^{178}$Hf  & -16 & \cite{10}  \\
3h & $^{248}$Cf & $^{16}$O+$^{232}$Th  & 24 & \cite{05} & 8c & $^{212}$Ra & $^{30}$Si+$^{182}$W  & -12 & \cite{19}  \\
4a & $^{202}$Pb & $^{18}$O+$^{184}$W  & -22 & \cite{15} & 8d & $^{214}$Ra & $^{30}$Si+$^{184}$W  & -10 & \cite{19}  \\
4b & $^{204}$Pb & $^{18}$O+$^{186}$W  & -20 & \cite{16} & 8e & $^{216}$Ra & $^{30}$Si+$^{186}$W  & -8 & \cite{19}  \\
4c & $^{210}$Po & $^{18}$O+$^{192}$Os  & -14 & \cite{08} & 8f & $^{227}$Np & $^{30}$Si+$^{197}$Au  & 5 & \cite{20}  \\
4d & $^{212}$Rn & $^{18}$O+$^{194}$Pt  & -12 & \cite{09} & 9a & $^{216}$Th & $^{32}$S+$^{184}$W  & -8 & \cite{13}  \\
4e & $^{216}$Rn & $^{18}$O+$^{198}$Pt  & -8 & \cite{09} & 9b & $^{230}$Pu & $^{32}$S+$^{198}$Pt  & 4 & \cite{23}  \\
5a & $^{197}$Tl & $^{19}$F+$^{178}$Hf  & -27 & \cite{12} & 10a & $^{208}$Rn & $^{48}$Ti+$^{160}$Gd  & -16 & \cite{10} \\
\end{tabular}
\end{ruledtabular}
\end{table*}
\begin{figure*}
\begin{center}
\includegraphics[width=18.0 cm, height=10.0 cm]{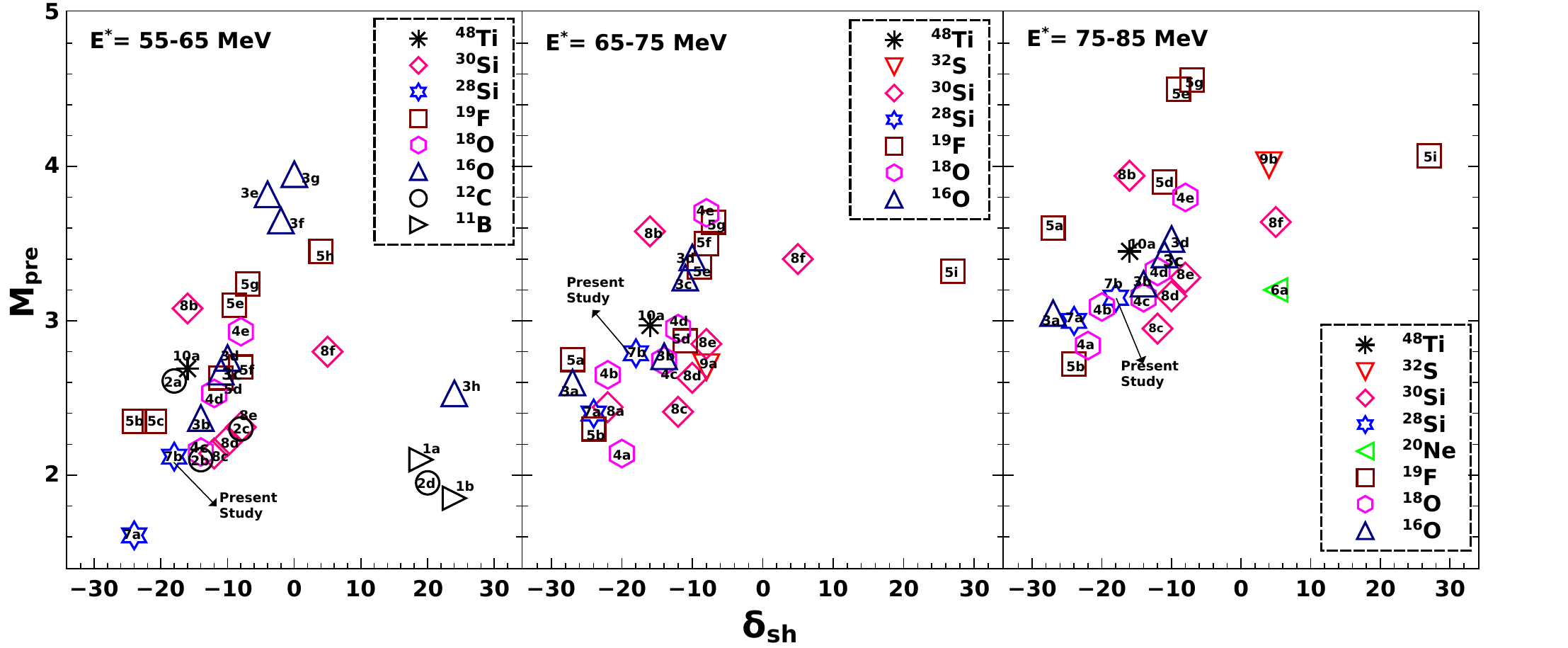}
\caption{Variation of available experimental data of M$_{pre}$, including present data (7b), with $\delta_{sh}$ in three different energy regions. The errors in M$_{pre}$ lies within the height of the symbols. Alpha-numeric value in each symbols identify the serial number in Table II.}
\end{center}
\end{figure*}
\begin{figure}
\begin{center}
\includegraphics[width=8.50 cm, height=12.0 cm]{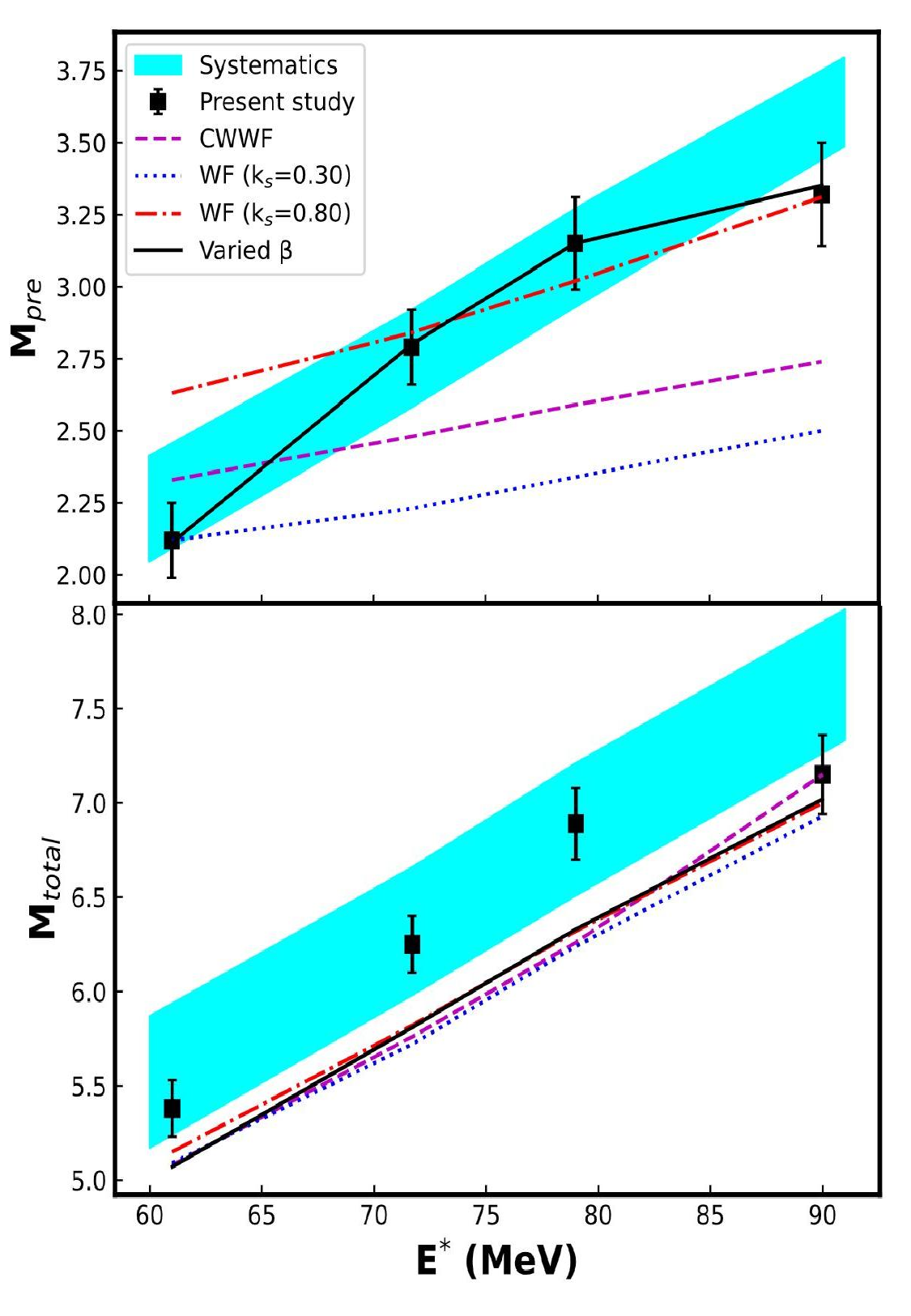}
\caption{Variation of experimentally and theoretically calculated M$_{pre}$ (top) and M$_{total}$ (bottom) w.r.t excitation energy for compound nucleus $^{206}$Rn.}
\end{center}
\end{figure}

We compare the available (including present measurements) M$_{pre}$ data for 40 reactions in the compound nuclear mass range of 197$\le$M$_{CN}$$\le$251. Details of the reactions are given in Table II. All the data are grouped into three different excitation energy ($E^{*}$) regions to minimize any dependence on the excitation energy. These regions are defined as 55 MeV $\le$ E$^{*}$ $\le$ 65MeV, 65 MeV $\le$ E$^{*}$ $\le$ 75 MeV, and 75 MeV $\le$ E$^{*}$ $\le$ 85MeV. Also, for a few reactions, there are multiple data points within the energy window of 10 MeV. Since M$_{pre}$ does not vary much within this energy range, we just pick a single point near the center of that region.\\
Here, $\delta_{sh}$ represents the deviation of the neutron and proton numbers in the target and projectile from the respective nearest magic numbers. For the chosen reactions, it is apparent that $^{16}O$ and $^{208}Pb$ are the doubly shell-closed reference nuclei to measure $\delta_{sh}$ of the projectile and target, respectively. For example, in our case of $^{28}$Si+$^{178}$Hf reaction,
\begin{equation}
  (\delta_{sh})_{proj, neut}= N_{^{28}Si}-N_{^{16}O} = 14-8 = 6  
\end{equation}
\begin{equation}
  (\delta_{sh})_{proj, prot}= Z_{^{28}Si}-Z_{^{16}O} = 14-8 = 6  
\end{equation}
 Similarly,
 \begin{equation}
  (\delta_{sh})_{targ, neut}= N_{^{178}Hf}-N_{^{208}Pb} = 106-126 = -20  
\end{equation}
\begin{equation}
  (\delta_{sh})_{targ, prot}= Z_{^{178}Hf}-Z_{^{208}Pb} = 72-82 = -10  
\end{equation}
Adding all four contributions, we get $\delta_{sh}$= -18. Hence, $\delta_{sh}$ is a cumulative measure of the deviation from the magicity of a given target-projectile combination. Here, we retain the sign of each term rather than use the absolute values of deviations, as an excess of particles from the magic number may exert distinct effects than a shortfall of equivalent magnitude. The variation of M$_{pre}$ with $\delta_{sh}$ is illustrated in Fig. 5, for three energies group mentioned above. A systematic behaviour with a maximum at $\delta_{sh}$ = 0 and a nearly symmetric sharp fall on either side is quite visible in the energy region 55-65 MeV. Our experimental data (7b) also following the trend of this systematic. Scarcity of data in the $\delta_{sh}$ $>$ 0 region, leads us to more experimental investigation in the othe two energy regions.
\subsection{Dynamical model calculations}
The dynamics of the compound nuclei is analyzed utilizing the computational code VECLAN, which is centered on the one-dimensional Langevin dynamical model governed by the Langevin equations.  Additional information regarding this program is provided in Ref. \cite{41}.\\
In this study, we measured the neutrons emitted from each fissioning nucleus and its corresponding fragments. For each beam energy, averaging is performed over an ensemble of 10$^{6}$ Langevin events to estimate M$_{pre}$, M$_{post}$ and M$_{total}$.  For each event in the ensemble, the initial angular momentum is sampled according to the systematics outlined in Ref. \cite{42}.\\
Several prescriptions are used for the dissipation coefficient to replicate the experimental M$_{pre}$ and M$_{total}$.  Fig. 6 illustrates that prevalent options such as chaos-weighted wall friction (CWWF) \cite{43,44} and wall + window friction (WF) with a reduction factor k$_s$ = 0.30 do not accurately replicate the experimental M$_{pre}$ at elevated excitation energies, but they align well at lower excitation energy (specifically at 61.0 MeV) \cite{45}.  An improved agreement is observed when the wall + WF configuration is used with a reduction factor k$_s$ = 0.80, except for E$^{*}$=61 MeV. Furthermore, we performed computations using the shape-independent reduced dissipation ($\beta$) as a variable parameter.  Fig. 6 illustrates that variations in $\beta$ = 2.05, 4.65, 6.05 and 6.2 MeV/$\hbar$ align well with the experimental M$_{pre}$ results at E$^{*}$ of 61.0, 71.7, 79.0 and 90 MeV. Similarly, we have calculated M$_{total}$ for all the prescriptions given and observed that all the prescriptions give nearly similar results, which are slightly underpredicted from the experimental data. To ensure the precision of our data, we also calculated M$_{pre}$, and M$_{total}$ using systematic equations provided by Kozulin et al. \cite{46}.  Our calculated data match well with this systematic depicted by the shaded region in Fig. 6.\\ 

\subsection{Impact of dissipation and N/Z on neutron multiplicity around shell closure}
\begin{figure}
\begin{center}
\includegraphics[width=8.0 cm, height=7.0 cm]{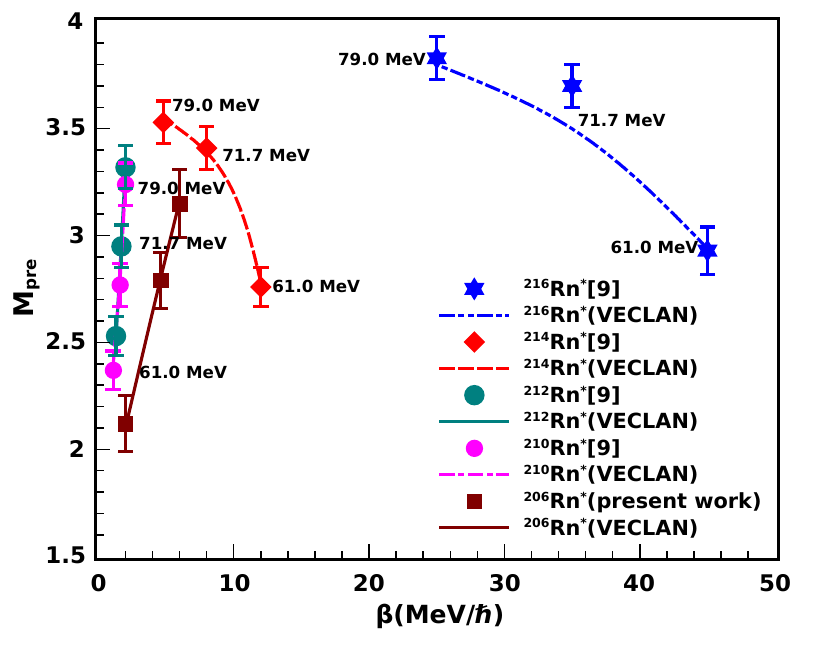}
\caption{Variation in $\beta$ w.r.t M$_{pre}$ for compound nucleus $^{206,210,212,214,216}$Rn with increase in excitation energy.}
\end{center}
\end{figure}
\begin{figure}
\begin{center}
\includegraphics[width=9.0 cm, height=15.0 cm]{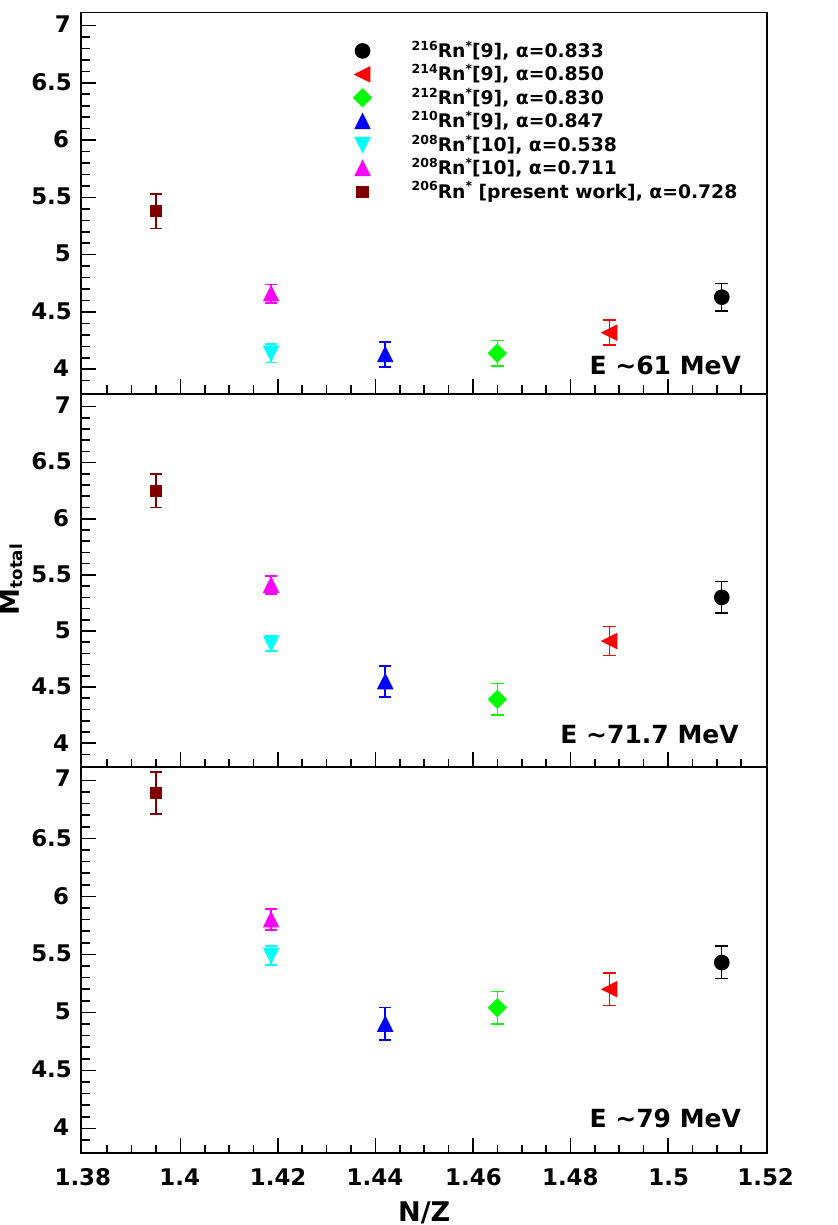}
\caption{Variation of M$_{total}$ w.r.t N/Z ratio for compound nucleus $^{206, 208, 210, 212, 214, 216}$Rn at excitation energies of 61.0 MeV, 71.7 MeV and 79.0 MeV.}
\end{center}
\end{figure}
To find the impact of dissipation we have used our present M$_{pre}$ measurement for the reaction $^{28}$Si+$^{178}$Hf forming CN $^{206}$Rn along with available experiment data of CN  $^{210,212,214,216}$Rn \cite{09}. We have calculated the variation in the dissipation parameter ($\beta$) for the experimental values M$_{pre}$ of the reactions mentioned above at three excitation energies i.e. 61.0, 71.7 and 79.0 MeV. The results of our calculations are plotted in Fig. 7, which illustrates that $\beta$ increases with rise in excitation energy for the compound nucleus $^{206}$Rn, which is below the shell closure (N=120). In contrast, for the compound nuclei $^{210,212}$Rn, situated near the shell closure (N=124,126), $\beta$ remains relatively constant as the excitation energies escalates.  However, once we surpass the shell-closure at CN $^{214,216}$Rn, a significant change in trend is observed; $\beta$ begins to decline with the increase in excitation energy. This indicates that shell closure region is very crucial to understand the dynamics of fission.\\
In this study, we also calculated the neutron-to-proton (N/Z) ratio in relation to the total neutron multiplicity (M$_{total}$). Our objective is to examine how the N/Z varies across a range of isotopes ($^{206,208,210,212,214,216}$Rn), with $^{206}$Rn being synthesized in the present work.
The CN $^{208}$Rn is formed using the reactions $^{30}$Si+$^{178}$Hf and $^{48}$Si+$^{160}$Gd \cite{10}, while $^{210,212,214,216}$Rn CN is formed from $^{16,18}$O+$^{194,198}$Pt \cite{09}. We compared the M$_{total}$ at three excitation energies: 61.0, 71.7 and 79.0 MeV, during our experiment we have matched these excitation energies with $^{210,212,214,216}$Rn and for $^{208}$Rn, we selected the closest available energy values. Fig. 8 presents the data that illustrate the correlation between M$_{total}$ and N/Z.  The figure further illustrates that neutron multiplicity escalates when one deviates from the shell closure, whether towards lower or higher neutron counts. However, at the shell closure, the neutron multiplicity diminishes, irrespective of the excitation energy. Furthermore, the reaction $^{48}$Si+$^{160}$Gd resulting in the compound nucleus $^{208}$Rn (down triangle) exhibits a reduced neutron multiplicity, this decrease is attributed to the occurrence of a quasi-fission contribution, which depends further on excitation energy \cite{10}.\\
\section{CONCLUSION} 
In this study, we analyzed the effect of dissipation near the neutron shell closure. Our findings indicate that below the shell closure, dissipation in the compound nucleus increases with rising excitation energy. At the shell closure, dissipation remains stable, while beyond the shell closure, it gradually decreases as excitation energy increases. Furthermore, we investigated the variation of the neutron-to-proton (N/Z) ratio with total neutron multiplicity at different excitation energies for $^{206,208,210,212,214,216}$Rn compound nuclei. Our analysis revealed a novel trend; near the shell closure CN the total neutron multiplicity decreases, but beyond the shell closure, it begins to increase either lower side or higher side.
\begin{acknowledgments}
The authors acknowledge the Pelletron+LINAC group and Target Laboratory of IUAC for their support during the experiment. One of the authors (Punit Dubey) is grateful to the Prime Minister Research Fellowship (PMRF) for the financial support for this work. One of the authors (Ajay Kumar) would like to thank the Institutions of Eminence (IoE) BHU [Grant No. 6031-B] and  IUAC-UGC, Government of India (Sanction No. IUAC/XIII.7/UFR-71353). The authors acknowledge the National Supercomputing Mission (NSM) for providing resources of PARAM Shivay at Indian Institute of Technology (BHU), Varanasi, which is implemented by C-DAC and supported by the Ministry of Electronics and Information Technology (MeitY) and Department of Science and Technology (DST), Government of India.
\end{acknowledgments}
\bibliography{basename of .bib file}

\end{document}